\documentclass{appolb}
\usepackage{epsfig}
\begin{document}
\title{Neutrino events from SN1987A revisited%
\thanks{Presented at 29th Mazurian Lakes Conference on Physics,
  Piaski, Poland, 2005.}%
\hspace{1mm}
\thanks{This Work has been partly supported by the Polish Ministry of
  Scientific Research and Information Technology, grants 1 P03B 049 26
  and 160/E-340/SPB/ICARUS/P-03/DZ 212/2003-2005.}%
}
\author{B.Bekman$^{1)}$, J.Holeczek$^{2)}$, J.Kisiel$^{2)}$
  \address{$^{1)}$ M. Sk\l{}odowska-Curie Memorial Cancer Center and
    Institute of Oncology, Gliwice Branch , ul. Wybrze\.za Armii
    Krajowej 15, 44-101 Gliwice, Poland.\\
    $^{2)}$ Institute of Physics, University of Silesia,
    ul. Uniwersytecka 4, 40-007 Katowice, Poland.}
}
\maketitle
\begin{abstract}
The $\e^{-}$ and $\e^{+}$ energy spectra from the SN1987A Supernova
neu\-tri\-no burst interactions are calculated and compared to the
observed spectra in Kamiokande-II and IMB experiments.
Neu\-tri\-no oscillations in Supernova and regeneration effects in the
Earth, for four combinations of neu\-tri\-no mass hierarchy
(Direct/Inverted) and the value of the mixing angle $\theta_{13}$
(Large/Small), are taken into account.
The influence of the (an\-ti)neu\-tri\-no production spectra
in Supernova on the observed, in Kamiokande-II and IMB detectors,
$\e^{-}$/$\e^{+}$ spectra is shown.
\end{abstract}
%
%===================================================================
\section{Introduction}
%===================================================================
%
When a thermal pressure of the heat, resulting from the sequence of
nuclear fusion reactions taking place in a star, cannot longer
balance the gravitational attraction, the last stage of a massive star
evolution begins.
For sufficiently massive stars (mass $>10M$$_{\odot }$), the collapse
of their iron core with a mass exceeding the Chandrasekhar limit
creates a type II Supernova (SN), which explosion is an intense source
of neu\-tri\-nos. It is estimated that, about $\sim $99\% of the SN type
II binding energy (about $\sim $10$^{53}$ erg) is radiated as
(an\-ti)neu\-tri\-nos of all flavors.
The SN neu\-tri\-no flux is so enormous, that even (an\-ti)neu\-tri\-no signals
from SN explosions outside our Galaxy can be detected on the Earth.
Up to now, there was only one case of a detection of extra-solar
system neu\-tri\-nos, namely neu\-tri\-nos from the SN1987A, which had
exploded in the Large Magellanic Cloud.
Two terrestrial detectors, Kamiokande-II~\cite{kamiokande} and
IMB~\cite{imb}, were able to observe 20 events of the SN1987A
neu\-tri\-no interactions in total.
This observation confirms the main features of the mechanism of the
stellar evolution.
%
%===================================================================
\section{The Supernova 1987A Signal}
%===================================================================
%
The number of observed neu\-tri\-no events, in the terrestrial detectors,
and their energy spectrum depend on:
(1) the assumed neu\-tri\-no production fluxes,
(2) the changes in neu\-tri\-no fluxes due to the neu\-tri\-no flavor mixing
(oscillations) in both, the Supernova and in the Earth, and
(3) the detector properties - including detection efficiency and
the dead time.

All these effects were taken into account in our calculations of the
electron and positron energy spectra from the SN1987A neu\-tri\-no
interactions in Kamiokande-II and IMB detectors (these detectors were
not able to distinguish between electrons and positrons), presented in
this paper.

For the purpose of our simulations, we assume that the distance
between the SN1987A and the Earth is 52kpc and that the total energy
emitted in the core collapse of the star in form of (an\-ti)neu\-tri\-nos is
$L_{tot} \simeq 3 \times 10^{53} ergs$.
%
%===================================================================
\subsection{Primary Spectra}
%===================================================================
%
We focus on the spectral characteristics of the
time-integrated (an\-ti)neu\-tri\-nos fluxes.
The detailed spectral shape of the production fluences is not well
known.
Such fluxes are often expressed in the form of Fermi-Dirac
distributions with the chemical potential equal to zero (see
also~\cite{Bekman}),
\begin{eqnarray*}
\mathop{\mathrm{F^0_\alpha (E)}}=
\frac
{\displaystyle 120}
{\displaystyle 7 {\pi}^4}
\cdot \frac
{\displaystyle L_\alpha}
{\displaystyle T^4_\alpha}
\cdot \frac
{\displaystyle E^2}
{\displaystyle e^{E\slash {T \alpha}} + 1},
\end{eqnarray*}
where $\alpha =\nu_{e}$, $\nu_{\mu }$, $\nu_{\tau }$, $\bar \nu_{e}$,
$\bar \nu_{\mu}$, $\bar \nu_{\tau}$ and $E$ represents the energy of
(an\-ti)neu\-tri\-nos.
The $L_\alpha$ is the total energy released in various flavors of
(an\-ti)neu\-tri\-nos ($L_\alpha \simeq L_{tot}/6$,
where an equipartition of the energy emitted in different species
is assumed), the $T_\alpha$ is the temperature of the $\nu_\alpha$ gas
in the (an\-ti)neu\-tri\-no sphere.
In our calculations, we assume the following, "standard", hierarchy of
temperatures: $T_{\nu_e} = 3.5 MeV$, $T_{\bar {\nu}_{e}} = 5 MeV$,
$T_{\nu_x,\bar{\nu}_x} = 8 MeV$, where $\nu_x$ and  $\bar{\nu}_x$ mean
$\nu_{\mu}$, $\nu_{\tau}$ and $\bar{\nu}_{\mu}$, $\bar{\nu}_{\tau}$,
respectively.
These temperatures corresponds to the following, approximate average
values of the neu\-tri\-no energies: $E_{\nu_e} = 11 MeV$,
$E_{\bar {\nu}_{e}} = 16 MeV$, $E_{\nu_x,\bar{\nu}_x} = 25
MeV$.

Using a nominal Fermi-Dirac distribution to approximate the spectrum is
physically motivated because a truly thermal neu\-tri\-no flux would
follow this behavior.
Numerical simulations of the Supernova explosion suggest, however,
that the production fluences can be "pinched", meaning that their low-
and high-energy parts are suppressed relative to the Fermi-Dirac
spectrum of the same average energy.
Therefore, we consider also an alternative "Analytic Fit
Function"~\cite{Keil} for which analytic simplicity is the main
motivation,
\begin{eqnarray*}
\mathop{\mathrm{F^0_\alpha (E)}}=
\frac
{\displaystyle L_{\alpha}}
{\displaystyle E^2_{\alpha}}
\cdot \frac
{\displaystyle (\beta_{\alpha}+1)^{{\beta \alpha}+1}}
{\displaystyle \Gamma (\beta_{\alpha}+1)}
\cdot \bigg (\frac
{\displaystyle E}
{\displaystyle E_{\alpha}}\bigg )^{\beta \alpha}
\cdot {\displaystyle e^{-({\beta \alpha}+1) E\slash {E \alpha}}},
\end{eqnarray*}
where $E$ is the energy of (an\-ti)neu\-tri\-nos, $L_{\alpha}$ is the
total energy released, $E_{\alpha}$ is the average energy and
$\beta_{\alpha}$ is the amount of spectral pinching.

In our analysis, we sum up the spectra coming from the two phases of
the Supernova explosion, the "accretion" phase in which (about) 10\%
of the total $L_{tot}$ is released, and the "cooling" phase in which the
remaining 90\% of the $L_{tot}$ is released.
The values of parameters, that we use, were extracted from the Garching
group paper~\cite{Raffelt} and are given in Table~\ref{tab.Keil}.
\begin{table}\centering
\begin{tabular}{|l||c|c|c|c|c|c|c|c|}
\hline
% & & & & & & & &  \\
\ & $E_{\nu_{e}} $ & $E_{\bar {\nu}_{e}} $ &
$E_{\nu_{x},\bar{\nu}_x} $ & $ \beta_{\nu_{e}}$& $\beta_{\bar {\nu}_{e}}$&
$\beta_{\nu_{x},\bar{\nu}_x}$& $\frac{{L \nu_{e}}}{{L \nu_{x},\bar{\nu}_x}}$ & $\frac{{L
    \bar{\nu}_{e}}}{{L \nu_{x},\bar{\nu}_x}}$ \\
% & & & & & & & &  \\
\hline
Accretion & 11.0 & 13.5 & 15.0 & 4.0 & 4.5 & 3.0 & 1.4 & 1.4 \\
\hline
Cooling & 13.5 & 16.5 & 18.0 & 2.7 & 2.7 & 2.7 & 0.7 & 0.7 \\
\hline  
\end{tabular}
\caption[]{The values of the parameters, of the "Analytic Fit
  Functions", used in this paper. For details see text
  and~\cite{Raffelt}.}
\label{tab.Keil}
\end{table}
%
%===================================================================
\subsection{Modifications by Oscillations}
%===================================================================
%
The details of our approach can be found in~\cite{Bekman}, so here
we only recall the main features:
(1) neu\-tri\-no oscillations in SN are expressed in terms of $\it flip$
$\it probabilities$ describing their transitions in two resonance
layers~\cite{Dighe},
(2) due to the large distance from the Supernova, they reach the Earth
surface in form of incoherent mass eigenstates,
(3) for the neu\-tri\-no regeneration in the Earth, the realistic
PREM~I~\cite{Dziewonski} Earth density profile is used.

We consider two neu\-tri\-no mass hierarchies - the Direct (Normal, $m_1 <
m_2 << m_3$) and Inverted ($m_3 << m_1 < m_2$) ones, and two values of
$\Theta_{13}$ (called "Large" and "Small",
see also~\cite{Bekman}).
The following values of the neu\-tri\-no mixing parameters were
used (\cite{Fogli}, the Dirac's phase $\delta = 0$):
\begin{eqnarray*}
\Delta m^2_{21} & = & 7.92 \times 10^{-5} eV^2\\
\Delta m^2_{32} & = & 2.4 \times 10^{-3} eV^2\\
\sin^2{\Theta_{12}}& = &0.314\\
\sin^2{\Theta_{23}}& = &0.44\\
\sin^2{\Theta_{13}}& = &0.9\times10^{-2} \hspace{1.9cm} Large \enspace \Theta_{13}\\
\sin^2{\Theta_{13}}& = &1.0\times10^{-7} \hspace{1.9cm} Small \enspace \Theta_{13}\\
\end{eqnarray*}
%
%===================================================================
\subsection{Kamiokande-II and IMB detectors}
%===================================================================
%
Both these detectors are water \v Cerenkov detectors.
For the description of the Kamiokande-II detector we
use~\cite{kamiokande} (2.14ktons of water, the efficiency extracted
from the FIG.3 in there, dead time negligible),
while for the description of the IMB detector we use~\cite{imb}
(6.8ktons of water, the efficiency extracted from the FIG.1 in there,
13\% dead time correction).

The Charge Current (CC) (an\-ti)neu\-tri\-no interactions with water, which
are considered in our calculations
(together with their calculated percentage contributions to the total
number of interactions in these detectors),
are given in Table~\ref{tab.CC}
(the cross sections for all processes come from~\cite{Heise}).
\begin{table}\centering
\begin{tabular}{|l||c|c|}
\hline
 &Fermi-Dirac&Analytic Fit Function\\
\hline
$\bar \nu _{e}+p\rightarrow n+e^+$ & 76$\%$-80$\%$& 87$\%$-89.5$\%$\\
\hline
$\nu _{e}+O\rightarrow F+e^-$ &6$\%$-10$\%$ & 2$\%$-3.5$\%$\\
\hline
$\bar {\nu} _{e}+O\rightarrow N+e^+$ &2.5$\%$-10$\%$ & 1.5$\%$-2$\%$ \\
\hline
\end{tabular}
\caption[]{The main CC (an\-ti)neu\-tri\-no interactions with water,
  together with their calculated percentage contributions to the total
  number of interactions (given ranges correspond to different
  combinations of neu\-tri\-no mass hierarchy and $\Theta_{13}$ value).}
\label{tab.CC}
\end{table}

\begin{figure}[ht]
\begin{center}
\mbox{\epsfig{figure=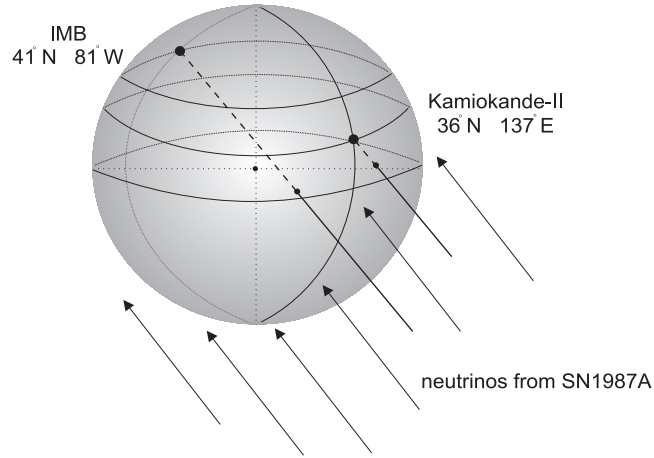,width=8.5cm,height=6.0cm}}
\caption[]{The sketch of the positions of Kamiokande-II and IMB
  detectors at the time of the arrival of SN1987A neu\-tri\-nos.}
\label{fig.crossing_earth}
\end{center}
\end{figure}

The Fig.\ref{fig.crossing_earth} shows the sketch of the positions of
Kamiokande-II and IMB detectors at the time of the arrival of SN1987A
neu\-tri\-nos.
We assume the following distances traveled by (an\-ti)neu\-tri\-nos in the
Earth during the SN1987A neu\-tri\-no burst (see also~\cite{Lunardini}):
4363km for Kamiokande-II and 8535km for IMB.
%
%===================================================================
\section{Results and discussion}
\begin{figure}
\begin{center}
\includegraphics [width=12.5cm,height=5.6cm]{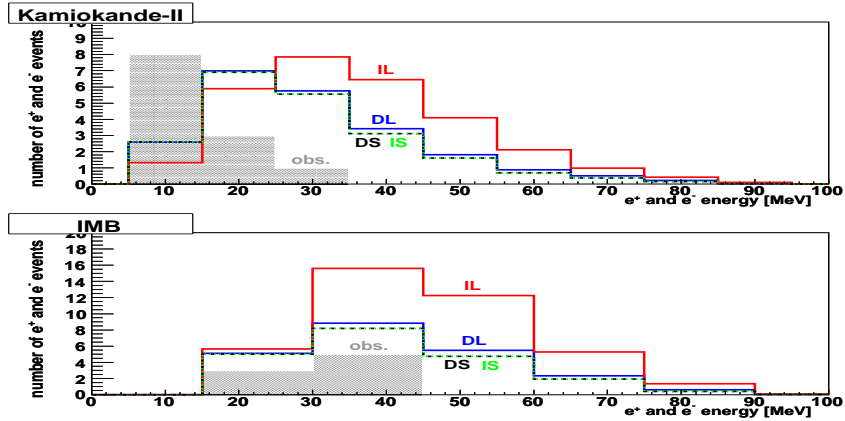}
\end{center}
\caption[]{The predicted $e^{+}$ and $e^{-}$ energy spectra,
         in case the SN1987A production spectra are described by the thermal
         Fermi-Dirac functions, for different combinations
         of mass hierarchy and $\Theta_{13}$ (DL - Direct mass
         hierarchy and $Large \enspace \Theta_{13}$, IL - Inverted
         mass hierarchy and $Large \enspace \Theta_{13}$, DS - Direct
         mass hierarchy and $Small \enspace \Theta_{13}$, IS -
         Inverted mass hierarchy and $Small \enspace
         \Theta_{13}$). The shaded areas show the histograms of
         observed SN1987A events. For details see the description in text.}
\label{fig.SN0_K2_IMB}
\end{figure}

\begin{figure}
\begin{center}
\includegraphics [width=12.5cm,height=5.6cm]{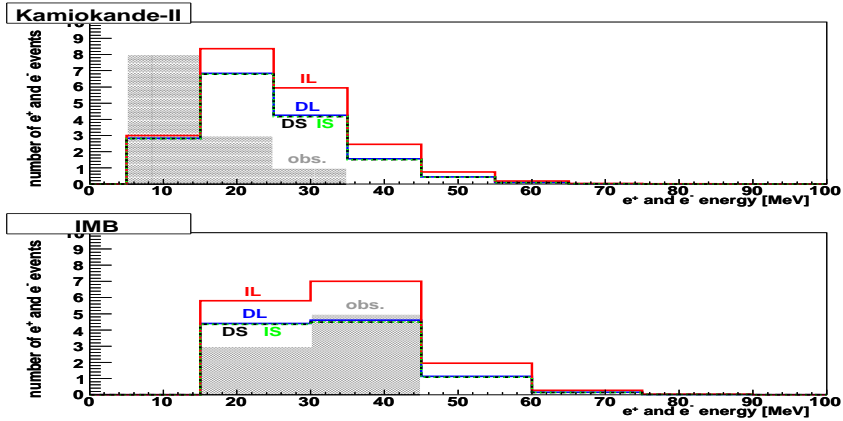}
\end{center}
\caption[]{The predicted $e^{+}$ and $e^{-}$ energy spectra,
         in case the SN1987A production spectra are described by the "Analytic
         Fit Functions", for different combinations
         of mass hierarchy and $\Theta_{13}$.
	 For details see the description in Fig.\ref{fig.SN0_K2_IMB}
         and in text.}
\label{fig.SN17_K2_IMB}
\end{figure}

%===================================================================
%
The predicted $e^{+}$ and $e^{-}$ energy spectra, from the SN1987A
neu\-tri\-no interactions, in Kamiokande-II and IMB detectors are
presented in Figs.~\ref{fig.SN0_K2_IMB} and~\ref{fig.SN17_K2_IMB},
respectively, for the Fermi-Dirac and the "Analytic Fit Functions"
describing the Supernova neu\-tri\-no production spectra.

The lines correspond to the four combinations of neu\-tri\-no mass
hierarchy and $\Theta_{13}$ value, namely DL - Direct mass hierarchy
and $Large \enspace \Theta_{13}$, IL - Inverted mass hierarchy and
$Large \enspace \Theta_{13}$, DS - Direct mass hierarchy and $Small
\enspace \Theta_{13}$, IS - Inverted mass hierarchy and $Small
\enspace \Theta_{13}$.
The shaded areas show the histograms of the observed SN1987A neu\-tri\-no
events, respectively, 12 and 8, in the Kamiokande-II and IMB detectors.

The predicted total numbers of registered events, in both detectors,
are given in Table~\ref{tab.N}.
The numbers in parenthesis give the predicted total numbers of
neu\-tri\-no interactions $N$ in these detectors (all possible reactions
taken into account, no detector efficiency and no dead time
considered, see also~\cite{Bekman}).
\begin{table}\centering
\begin{tabular}{|l||c|c|c|c||c|c|c|c|}
\hline
 &DL &DS &IL &IS  \\
\hline
\hline
 & \multicolumn{4}{|c||}{Fermi-Dirac} \\
\hline
 Kamiokande-II&22.2&21.0&29.3&21.0\\
 &(30.1)&(29.4)&(36.5)&(29.4)\\
\hline
 IMB&22.4&20.3&40.2&20.3\\
 &(95.6)&(93.3)&(116.1)&(93.3)\\
\hline
\hline
 & \multicolumn{4}{|c|}{Analytic Fit Function} \\
\hline
Kamiokande-II&16.0&15.8&20.7&15.8\\
&(21.0)&(20.8)&(25.9)&(20.8)\\
\hline
IMB&10.3&10.1&15.0&10.1\\
&(66.8)&(66.1)&(82.5)&(66.1)\\
\hline
\end{tabular}
\caption[]{The predicted total numbers of registered neu\-tri\-no events
  in Kamiokande-II and IMB detectors. The numbers in parenthesis give
  the predicted total numbers of neu\-tri\-no interactions $N$ in these
  detectors. For details see text.}
\label{tab.N}
\end{table}

It can be seen that none of the simulated spectra, obtained with any of
the considered neu\-tri\-no production energy spectra, and any of the mass
hierarchy and $\Theta_{13}$ combinations, agrees satisfactorily with the
measured distributions.
The agreement is slightly better for the IMB spectra, especially for
the "Analytic Fit Function" neu\-tri\-no production spectra.

Moreover, with the current neu\-tri\-no mixing parameters, the different
Earth matter effects (i.e.~the different distances traveled by
neu\-tri\-nos in the Earth) cannot also explain the qualitative
differences in the observed Kamiokande-II and IMB spectra.
In fact, as we have already noticed in~\cite{Bekman}, the Earth
matter effects have almost no influence on the expected total number
of neu\-tri\-no interactions $N$ in detectors, nor on the predicted
event spectra, except on distances below about 2500km, where some
variations within 3.5\% (1.5\%), for the Fermi-Dirac ("Analytic Fit
Function") neu\-tri\-no production spectrum exist (one gets the highest
numbers of events when neu\-tri\-nos do not cross the Earth at all).

The shape of all four (DL, IL, DS and IS) simulated $e^{+}$ and
$e^{-}$ energy spectra is very similar.
However, the number of the expected events is considerably higher (and
the spectrum is harder) for the IL case than for the remaining three,
for which the distributions are practically identical (note also
that, all these numbers scale linearly with the assumed total energy
$L_{tot}$ released in form of (an\-ti)neu\-tri\-nos, and are inversely
proportional to the square of the assumed distance to the Supernova).

The next generations of neu\-tri\-no experiments, with much bigger
detectors, and improved detection techniques, should be able to
observe future Supernova neu\-tri\-no bursts, and not only confirm the
main features of the SN mechanism, but also to allow for more
precise tests of SN models.
\end{document}